\documentstyle[epsfig]{preprint}

\newcommand{\fract}[2]{{\textstyle\frac{#1}{#2}}}

\newcommand{\lapeq}{\stackrel{\scriptscriptstyle\raisebox{-3.0mm}{$<$}}
{\scriptscriptstyle \raisebox{-1.5mm}{$\sim$}}}

\begin{document}
\title{STRONGLY DISTORTED BARYON WAVE--FUNCTIONS: HYPERON
BETA--DECAY AND THE SPIN OF THE $\Lambda$ AND THE
NUCLEON\footnote{Talk presented at the international conference
{\it Symmetry and Spin} Prague, July 2000. To appear in the Proceedings.}
}
\authori{Herbert Weigel}
\addressi{Center for Theoretical Physics, Lab for
Nuclear Science and Dept of Physics \\
Massachusetts Institute of Technology, Cambridge, Massachusetts 02139}
\authorii{}     
\addressii{Their Affiliation and Address}
\authoriii{}     
\addressiii{Their Affiliation and Address}
\headtitle{STRONGLY DISTORTED BARYON WAVE--FUNCTIONS \ldots}
\headauthor{H. Weigel}  
\specialhead{H. Weigel: STRONGLY DISTORTED BARYON WAVE--FUNCTIONS \ldots}
\evidence{A}
\daterec{}    
\cislo{0}  \year{2000}
\setcounter{page}{1}
\pagesfromto{000--000}
\maketitle

\begin{abstract}

Within the collective coordinate approach to chiral soliton
models we suggest that breaking of $SU(3)$ flavor symmetry mainly
resides in the baryon wave--functions while the charge operators
have no (or only small) symmetry breaking components. In this 
framework we study the $g_A/g_V$ ratios for hyperon beta--decay as 
well as the various quark flavor components of the axial charge of 
the nucleon and the $\Lambda$--hyperon. 

\bigskip
\noindent
MIT--CTP--3017
\end{abstract}

\section{Introduction and Motivation}

Using results on the axial current matrix elements from deep--inelastic
scattering as well as hyperon beta--decay data together with flavor 
covariance results in sizable polarizations for the {\it non--strange} 
quarks, $\Delta U_\Lambda=\Delta D_\Lambda\approx-0.20$ together with 
$\Delta S_\Lambda\approx0.60$ for the {\it strange} quark inside the 
$\Lambda$--hyperon~\cite{Ja96}. The assumption of flavor covariance is 
motivated by the feature that the 
Cabibbo scheme \cite{Ca63} utilizing the $F$\&$D$ parameterization for 
the flavor changing axial charges works unexpectedly well~\cite{Fl98} as 
the comparison in table~\ref{empirical} exemplifies. 

\begin{table}[hb]
~\vskip-0.5cm
\caption{\label{empirical}\sf The empirical values for the 
$g_A/g_V$ ratios of hyperon beta--decays \protect\cite{DATA}, 
see also~\protect\cite{Fl98}. 
For the process $\Sigma\to\Lambda$ only $g_A$ is given.
Also the flavor symmetric predictions are presented using the 
values for $F$\&$D$ which are mentioned in section III. Analytic
expressions which relate these parameters to the $g_A/g_V$ ratios
may {\it e.g.} be found in table I of \protect\cite{Pa90}.}
~\vskip0.01cm
{\small
\begin{tabular}{ c || c | c | c | c | c}
& $\Lambda\to p$ & $\Sigma\to n$ & $\Xi\to\Lambda$ &
$\Xi\to\Sigma$ & $\Sigma\to\Lambda$\\
\hline
emp.& $0.718\pm0.015$ & $0.340\pm0.017$ & $0.25\pm0.05$ &
$1.287\pm0.158$ & $0.61\pm 0.02$\\
$F$\&$D$& $0.725\pm0.009$ & $0.339\pm0.026$ & $0.19\pm0.02$ &
$1.258=g_A$ & $0.65\pm0.01$
\end{tabular}}
\end{table}

To account for flavor symmetry breaking effects we consider the Skyrme 
model approach in which baryons emerge as solitons in an effective 
meson theory. In such models baryon states are obtained by quantizing 
the large amplitude fluctuations (zero modes) about the soliton. 
Exact eigenstates are obtained for any strength of symmetry 
breaking~\cite{Ya88}. We focus on a picture with the symmetry breaking 
mainly residing in the baryon wave--functions, 
including important contributions which would be missed in a first
order treatment. In contrast, we assume that the current 
operators, from which the charges are computed, are dominated by their 
flavor covariant components. This approach approximately reproduces 
the data with no (or only minor) explicit symmetry breaking in the 
charge operators. In addition we present the results obtained from a 
realistic vector meson soliton model that supports the suggested 
picture. Details omitted here may be traced from ref~\cite{We00}.

\section{Symmetry Breaking in the Baryon Wave--Functions} 

Here we review the collective coordinate quantization for the 
low--lying $\frac{1}{2}^+$ and $\frac{3}{2}^+$ baryons in soliton models. 
The collective coordinates $A$ are introduced via
\be
U(\vec{r},t)=A(t)U_0(\vec{r})A^\dagger(t)\, ,\qquad 
A(t)\in SU(3)\, .
\label{collcord}
\ee
$U_0(\vec{r})$ describes the soliton embedded
in the isospin subgroup. A prototype model Largangian
for the chiral field $U(\vec{r},t)$ would consist of the Skyrme model 
supplemented by the Wess--Zumino--Witten term as well as suitable 
symmetry breaking pieces. We parameterize the collective coordinates 
by eight ``Euler--angles'' 
\be
A=D_2(\hat{I})\,{\rm e}^{-i\nu\lambda_4}D_2(\hat{R})\,
{\rm e}^{-i(\rho/\sqrt{3})\lambda_8}\ ,
\label{Apara}
\ee
where $D_2$ denote rotation matrices of three Euler--angles for 
each, rotations in isospace~($\hat{I}$) and 
coordinate--space~($\hat{R}$). Substituting (\ref{collcord})
into the model Lagrangian yields 
upon canonical quantization the Hamiltonian for the
collective coordinates~$A$: 
\be 
H=H_{\rm s}+\fract{3}{4}\, \gamma\, {\rm sin}^2\nu\, .
\label{Hskyrme}
\ee
The symmetric piece of this collective Hamiltonian only contains 
Casimir operators and may be expressed in terms of the $SU(3)$--right 
generators $R_a\, (a=1,\ldots,8)$:
\be
H_{\rm s}=M_{\rm cl}+\frac{1}{2\alpha^2}\sum_{i=1}^3 R_i^2
+\frac{1}{2\beta^2}\sum_{\alpha=4}^7 R_\alpha^2\, .
\label{Hsym}
\ee
$M_{\rm cl},\alpha^2,\beta^2$ and $\gamma$ are functionals of the 
soliton, $U_0(\vec{r})$. The generators $R_a$ can be expressed in terms 
of derivatives with respect to the `Euler--angles'. The eigenvalue 
problem $H\Psi=\epsilon\Psi$ reduces to sets of ordinary second order 
differential equations for isoscalar functions which only depend on 
the strangeness changing angle $\nu$~\cite{Ya88}. Only the product 
$\omega^2=\frac{3}{2}\gamma\beta^2$ appears in these differential 
equations which is thus interpreted as the effective strength of the 
flavor symmetry breaking. A value in the range $5\lapeq\omega^2\lapeq8$ 
is required to obtain reasonable agreement with the empirical mass 
differences for the $\frac{1}{2}^+$ and $\frac{3}{2}^+$ 
baryons~\cite{We96}. 

\section{Charge Operators} 

In the soliton description the effect of the derivative type symmetry 
breaking terms is mainly indirect. They provide the splitting between the 
various decay constants and thus increase $\gamma$ which is proportional 
to $f_K^2m_K^2-f_\pi^2m_\pi^2\approx 1.5f_\pi^2(m_K^2-m_\pi^2)$. 
Otherwise the derivative type symmetry breaking 
terms may be omitted. Whence there are no symmetry
breaking terms in current operators and the non--singlet axial charge 
operator is parameterized as ($ a=1,\ldots,8,\, i=1,2,3$)
\be
\int d^3r A_i^{(a)} = c_1 D_{ai} - c_2 D_{a8}R_i
+c_3\sum_{\alpha,\beta=4}^7d_{i\alpha\beta}D_{a\alpha}R_\beta
\, ,
\label{axsym}
\ee
where
$D_{ab}=\frac{1}{2}{\rm tr}\left(\lambda_a A\lambda_b A^\dagger\right)$.
In the limit $\omega^2\to\infty$
(integrating out {\it strange} degrees of freedom)
the strangeness contribution to the axial charge
of the nucleon should vanish. Noting that
$\langle N| D_{83}| N\rangle\to0$ and 
$\langle N| \sum_{\alpha,\beta=4}^7d_{3\alpha\beta}D_{8\alpha}R_\beta
| N\rangle\to0$ while $\langle N| D_{88}| N\rangle\to1$ for 
$\omega^2\to\infty$, we demand 
\be
\int d^3r A_i^{(0)}= -2\sqrt{3} c_2 R_i\quad i=1,2,3\,.
\label{singsym}
\ee
for the axial singlet current because it leads to the strangeness
projection, $A_i^{(s)}=(A_i^{(0)}-2\sqrt{3}A_i^{(8)})/3$ that vanishes
for $\omega^2\to\infty$.
Actually all model calculations in the literature \cite{Pa92,Bl93} 
are consistent with this requirement. 
In order to completely describe the hyperon beta--decays we
also demand matrix elements of the vector charges. These are obtained 
from the operator
\be
\int d^3r V_0^{(a)} = \sum_{b=1}^8D_{ab}R_b=L_a,
\label{vector}
\ee
which introduces the $SU(3)$--left generators $L_a$. 

\begin{figure}[t]
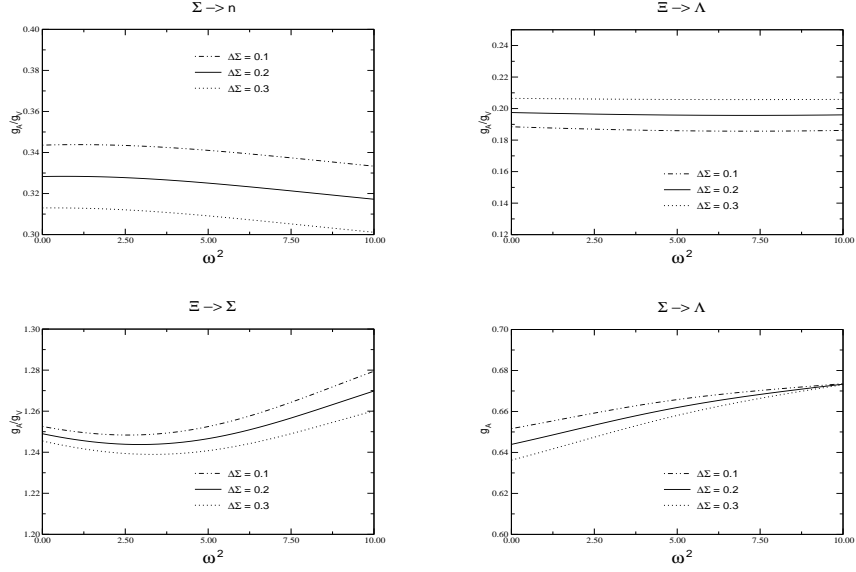

~\vskip0.05cm
\centerline{
\epsfig{figure=nusi.eps,height=5.0cm,width=3.5cm,angle=270}
\hspace{1.0cm}
\epsfig{figure=chla.eps,height=5.0cm,width=3.5cm,angle=270}}
~\vskip0.03cm
\centerline{
\epsfig{figure=sich.eps,height=5.0cm,width=3.5cm,angle=270}
\hspace{1.0cm}
\epsfig{figure=sila.eps,height=5.0cm,width=3.5cm,angle=270}}
\caption{\label{decay}\sf The predicted decay parameters for the 
hyperon beta--decays using $\omega^2_{\rm fix}=6.0$. 
The errors originating from those in $\Delta\Sigma_N$ are indicated.}
\end{figure}

The values for $g_A$ and $g_V$ 
(only $g_A$ for $\Sigma^+\to\Lambda e^+\nu_e$)
are obtained from the matrix elements of respectively 
the operators in eqs~(\ref{axsym}) and~(\ref{vector}), sandwiched 
between the eigenstates of the full Hamiltonian~(\ref{Hskyrme}).
We choose $c_2$ according the proton spin puzzle and subsequently 
determine $c_1$ and $c_3$ at $\omega^2_{\rm fix}=6.0$ such that the 
nucleon axial charge, $g_A$ and the $g_A/g_V$ ratio for 
$\Lambda\to p e^-\bar{\nu}_e$ are reproduced\footnote{In this section 
we will not address the problem of the too small model prediction for 
$g_A$ but rather use the empirical value $g_A=1.258$ as an input to 
determine the $c_n$.}. We are not only left with predictions for the 
other decay parameters but we can also study the variation with 
symmetry breaking. This is shown in figure~\ref{decay}.
The dependence on flavor symmetry breaking is very 
moderate\footnote{However, the individual matrix elements
entering the ratios $g_A/g_V$ vary strongly with 
$\omega^2$~\cite{We00}.} and the results can be viewed as reasonably 
agreeing with the empirical data, {\it cf.} table \ref{empirical}. 
The observed independence of $\omega^2$ shows that these
predictions are not sensitive to the choice of $\omega^2_{\rm fix}$.
We therefore have a two parameter ($c_1$ and $c_3$, $c_2$ is fixed 
from $\Delta\Sigma_N$) fit of the hyperon beta--decays. The two 
transitions, $n\to p$ and $\Lambda\to p$, which are not shown in 
figure~\ref{decay}, exhibit a similar neglegible dependence on $\omega^2$.
Comparing the results in figure \ref{decay} with the data in 
table~\ref{empirical} we see that the calculation using the strongly 
distorted wave--functions agrees equally well with the empirical 
data as the flavor symmetric $F$\&$D$ fit. We also observe that the 
singlet current does not get modified. Hence we have the simple relation
$\Delta\Sigma_N=\Delta\Sigma_\Lambda$ for all values of $\omega^2$.

\begin{figure}[t]
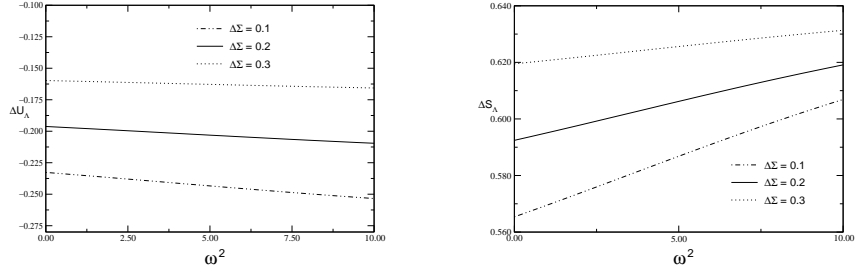

~\vskip-0.2cm
\centerline{
\epsfig{figure=h1la.eps,height=5.0cm,width=3.5cm,angle=270}
\hspace{1.0cm}
\epsfig{figure=h3la.eps,height=5.0cm,width=3.5cm,angle=270}}
\caption{\label{laxial}\sf The contributions of the {\it non--strange}
(left panel) and {\it strange} (right panel) degrees of freedom
to the axial charge of the $\Lambda$. Again we used
$\omega^2_{\rm fix}=6.0$.}
\end{figure}

In figure \ref{laxial} we display the flavor components of the axial 
charge of the $\Lambda$ hyperon. Again, the various contributions 
to the axial charge of the $\Lambda$ exhibit only a moderate dependence 
on $\omega^2$. The {\it non--strange} component, 
$\Delta U_\Lambda=\Delta D_\Lambda$ slightly increases in magnitude. 
The {\it strange} quark piece, $\Delta S_\Lambda$ grows with 
symmetry breaking since we keep $\Delta\Sigma_\Lambda$ fixed. Our
results agree nicely with an $SU(3)$ analysis applied to the 
data~\cite{Ja96} (see above). The observed
independence on the symmetry breaking does not occur for all
matrix elements of the axial current. An important counter--example
is the {\it strange} quark component in the nucleon, $\Delta S_N$. 
For $\Delta\Sigma=0.2$, say, it is significant at zero symmetry
breaking, $\Delta S_N=-0.131$ while it decreases (in magnitude) to 
$\Delta S_N=-0.085$ at $\omega^2=6.0$.

\section{Spin Content of the $\Lambda$ in a Realistic Model}

We consider a realistic soliton model containing pseudoscalar 
and vector meson fields. It has been established for two flavors in 
ref \cite{Ja88} and been extended to three flavors in ref~\cite{Pa92} 
where it has been shown to fairly describe the parameters of hyperon 
beta--decay ({\it cf.} table~4 in ref~\cite{Pa92}). 
The model Lagrangian contains terms which 
involve the Levi--Cevita tensor $\epsilon_{\mu\nu\rho\sigma}$, to 
accommodate processes like $\omega\rightarrow3\pi$~\cite{Ka84}. 
These terms contribute to $c_2$ and $c_3$.  A minimal set of 
symmetry breaking terms is included \cite{Ja89} to account 
for different masses and decay constants. These terms add 
symmetry breaking pieces to the axial charge operator,
\bea
\delta A_i^{(a)}=c_4 D_{a8}D_{8i}+
c_5 \sum_{\alpha,\beta=4}^7d_{i\alpha\beta}D_{a\alpha}D_{8\beta}+
c_6 D_{ai}(D_{88}-1)\,\, ,\,
\delta A_i^{(0)}= 2\sqrt{3}\,c_4D_{8i}\, .
\nonumber
\eea
The identical coefficient $c_4$ in the octet and singlet currents 
arises from the model calculation, it is not demanded by the 
consistency condition as $\omega^2\to\infty$.

Unfortunately the model parameters cannot be completely 
determined in the meson sector~\cite{Ja88}. We use the remaining 
freedom to accommodate baryon properties in three different ways as
shown in table \ref{realistic}. The set 
denoted by `b.f.' refers to a best fit to the baryon spectrum.
It predicts the axial charge somewhat on the low side, $g_A=0.88$.
The set named `mag.mom.' labels a set of parameters yielding 
magnetic moments close to the respective empirical data
(with $g_A=0.98$) and finally the set labeled `$g_A$' reproduces the 
axial charge of the nucleon and also reasonably accounts for hyperon
beta--decay \cite{Pa92}.
\begin{table}[htb]
\caption{\label{realistic}\sf Spin content of the $\Lambda$ in the
realistic vector meson model. For comparison the nucleon 
results are also given. Three sets of model parameters
are considered, see text.}
~\vskip0.01cm
{\small
\begin{tabular}{ c || c | c |c || c | c | c | c}
& \multicolumn{3}{c||}{$\Lambda$} &
\multicolumn{4}{c}{$N$}\\
\hline
& $\Delta U = \Delta D$ & $\Delta S$ & $\Delta\Sigma$ &
 $\Delta U$ & $\Delta D$ & $\Delta S$ & $\Delta\Sigma$\\
\hline
b.f.
&$-0.155$&$0.567$&$0.256$&$0.603$&$-0.279$&$-0.034$&$0.291$\\
mag. mom. 
&$-0.166$&$0.570$&$0.238$&$0.636$&$-0.341$&$-0.030$&$0.265$\\
$g_A$ 
&$-0.164$&$0.562$&$0.233$&$0.748$&$-0.476$&$-0.016$&$0.256$
\end{tabular}}
\end{table}
We observe that in particular the predictions for the axial properties 
of the $\Lambda$ are quite insensitive to the model parameters. Their
variation only influences the isovector part of the axial charge operator. 
The singlet matrix element of the $\Lambda$ hyperon is smaller than that 
of the nucleon. The full model calculation predicts sizable polarizations 
of the {\it up} and {\it down} quarks in the $\Lambda$ which are slightly 
smaller in magnitude but nevertheless comparable to those obtained 
from the $SU(3)$ symmetric analyses.

\section{Conclusions}

In the collective coordinate approach to chiral solitons large deviations 
from flavor symmetric (octet) wave--functions are required to accommodate 
the observed pattern of the baryon mass--splitting. We have 
suggested a picture for the axial charges of the 
low--lying~$\frac{1}{2}^+$~baryons which manages to reasonably reproduce 
the empirical data without introducing (significant) flavor symmetry 
breaking components in the corresponding operators. Rather, the sizable 
symmetry breaking resides almost completely in the baryon wave--functions. 
The empirical data for the parameters of hyperon beta--decay are as 
reasonably reproduced as in the Cabibbo scheme. We emphasize that the 
present picture is not a re--application of the Cabibbo scheme since in 
the present calculation the `octet' baryon wave--functions have significant 
admixture of higher dimensional representations. 

We may take the symmetry breaking parameter to be infinitely large.
For consistency then the two flavor model for the nucleon must be retrieved.
This consistency condition relates coefficients in the axial singlet 
current operator to the respective octet components. Disentangling the 
quark flavor components yields sizable {\it up} and {\it down} quark 
polarizations in the $\Lambda$. We also considered a realistic model, 
wherein the parameters entering the charge operators are actually 
predicted. This model calculation confirmed the results obtained in 
the parametrically treatment.

\bigskip
{\small 
The author is grateful to the organizers of the conference
for providing a fruitful and pleasant atmosphere. He also
thanks R.~L.~Jaffe and J.~Schechter
for helpful conversations and useful references.

This work is supported in part by funds provided by
the U.S. Department of Energy (D.O.E.) under cooperative research agreement
\#DF-FC02-94ER40818 and the Deutsche Forschungsgemeinschaft (DFG) under
contract We 1254/3-1.
}

\end {document}